\documentclass[aps, prl, reprint, superscriptaddress]{revtex4-1}
\usepackage[english]{babel}
\usepackage{amsmath,amsthm}
\usepackage{amsfonts}
\usepackage{xspace}
\usepackage{bm}
\usepackage[squaren, thinqspace]{SIunits}
\usepackage{booktabs}
\usepackage{graphicx}

\theoremstyle{definition}

\theoremstyle{remark}
\begin{document}
\renewcommand{\degree}{\ensuremath{^\circ}\xspace}
\renewcommand{\Re}{\ensuremath{\mathrm{Re}} \xspace}
\renewcommand{\Im}{\ensuremath{\mathrm{Im}} \xspace}
\newcommand{\Hy}{\ensuremath{\bm{H} || \bm{y}}\xspace}
\newcommand{\Hx}{\ensuremath{\bm{H} || \bm{x}}\xspace}
\newcommand{\y}{\ensuremath{\bm{y}}\xspace}
\newcommand{\x}{\ensuremath{\bm{x}}\xspace}
\newcommand{\z}{\ensuremath{\bm{z}}\xspace}
\newcommand{\Smag}{\ensuremath{\left|S_{21}\right|}\xspace}
\newcommand{\DSmag}{\ensuremath{\Delta\left|S_{21}\right|}\xspace}
\newcommand{\ii}{\ensuremath{\mathrm{i}}\xspace}
\newcommand{\LNO}{LiNbO$_3$\xspace}
\newcommand{\Pidt}{\ensuremath{P_\mathrm{IDT}}\xspace}
\newcommand{\Vdc}{\ensuremath{V_\mathrm{DC}}\xspace}
\newcommand{\DPidt}{\ensuremath{\Delta P_\mathrm{IDT}}\xspace}
\newcommand{\Pemw}{\ensuremath{P_\mathrm{EMW}}\xspace}
\newcommand{\Psaw}{\ensuremath{P_\mathrm{SAW}}\xspace}
\newcommand{\Vmsp}{\ensuremath{V_\mathrm{MSP}}\xspace}
\newcommand{\Vmr}{\ensuremath{V_\mathrm{MR}}\xspace}
\newcommand{\Vsp}{\ensuremath{V_\mathrm{SP}}\xspace}
\newcommand{\Vsse}{\ensuremath{V_\mathrm{SSE}}\xspace}
\newcommand{\bsp}{\ensuremath{b_\mathrm{SP}}\xspace}
\newcommand{\Esp}{\ensuremath{E_\mathrm{SP}}\xspace}
\newcommand{\DVdc}{\ensuremath{\Delta V_\mathrm{DC}}\xspace}
\newcommand{\DPsaw}{\ensuremath{\Delta P_\mathrm{SAW}}\xspace}
\newcommand{\DPemw}{\ensuremath{\Delta P_\mathrm{EMW}}\xspace}
\newcommand{\DVmsp}{\ensuremath{\Delta V_\mathrm{MSP}}\xspace}
\newcommand{\DVmr}{\ensuremath{\Delta V_\mathrm{MR}}\xspace}
\newcommand{\gSpinMix}{\ensuremath{g_{\uparrow\!\downarrow}}\xspace}
\newcommand{\VANE}{\ensuremath{V_\mathrm{ANE}}\xspace}
\newcommand{\VISHE}{\ensuremath{V_\mathrm{ISHE}}\xspace}
\newcommand{\Js}{\ensuremath{\bm{J}_\mathrm{s}}\xspace}
\newcommand{\mus}{\ensuremath{\bm{\mu}_\mathrm{s}}\xspace}
\newcommand{\Jc}{\ensuremath{\bm{J}_\mathrm{c}}\xspace}
\newcommand{\Jssp}{\ensuremath{J_\mathrm{s}^\mathrm{SP}}\xspace}
\newcommand{\Jssmr}{\ensuremath{J_\mathrm{s}^\mathrm{SMR}}\xspace}
\newcommand{\Jssse}{\ensuremath{J_\mathrm{s}^\mathrm{SSE}}\xspace}
\newcommand{\alphaSH}{\ensuremath{\alpha_{\mathrm{SH}}}\xspace}
\newcommand{\lambdaSD}{\ensuremath{\lambda_{\mathrm{SD}}}\xspace}
\newcommand{\tN}{\ensuremath{t_{\mathrm{N}}}\xspace}
\newcommand{\sigmaN}{\ensuremath{\sigma_{\mathrm{N}}}\xspace}
\newcommand{\tF}{\ensuremath{t_{\mathrm{F}}}\xspace}
\newcommand{\sigmaF}{\ensuremath{\sigma_{\mathrm{F}}}\xspace}
\newcommand{\J}{\ensuremath{\mathrm{J}}\xspace}

\title{Experimental test of the spin mixing interface conductivity concept}%

\author{Mathias Weiler}
\altaffiliation[present address: ]{National Institute of Standards and Technology, Boulder, CO, 80305}
\affiliation{Walther-Mei{\ss}ner-Institut, Bayerische Akademie der Wissenschaften, Garching, Germany}
\author{Matthias Althammer}
\author{Michael Schreier}
\affiliation{Walther-Mei{\ss}ner-Institut, Bayerische Akademie der Wissenschaften, Garching, Germany}
\author{Johannes Lotze}
\author{Matthias Pernpeintner}
\author{Sibylle Meyer}
\author{Hans Huebl}
\affiliation{Walther-Mei{\ss}ner-Institut, Bayerische Akademie der Wissenschaften, Garching, Germany}
\author{Rudolf Gross}
\affiliation{Walther-Mei{\ss}ner-Institut, Bayerische Akademie der Wissenschaften, Garching, Germany}
\affiliation{Physik-Department, Technische Universit\"{a}t M\"{u}nchen, Garching, Germany}
\author{Akashdeep Kamra}
\affiliation{Kavli Institute of Nanoscience, Delft University of Technology, Delft, The Netherlands}
\author{Jiang Xiao}
\affiliation{Department of Physics and State Key Laboratory of Surface Physics, Fudan University, Shanghai, China}
\author{Yan-Ting Chen}
\author{HuJun Jiao}
\affiliation{Kavli Institute of Nanoscience, Delft University of Technology, Delft, The Netherlands}
\author{Gerrit E. W. Bauer}
\affiliation{Kavli Institute of Nanoscience, Delft University of Technology, Delft, The Netherlands}
\affiliation{Institute of Materials Research  and WPI-AIMR, Tohoku University, Sendai, Japan}
\author{Sebastian T. B. Goennenwein}
\affiliation{Walther-Mei{\ss}ner-Institut, Bayerische Akademie der Wissenschaften, Garching, Germany}
\email{goennenwein@wmi.badw.de}
\date{\today}%
\begin{abstract}
  We perform a quantitative, comparative study of the spin pumping, spin Seebeck and spin Hall magnetoresistance effects, all detected via the inverse spin Hall effect in a series of over 20 yttrium iron garnet/Pt samples. Our experimental results fully support present, exclusively spin current-based, theoretical models using a single set of plausible parameters for spin mixing conductance, spin Hall angle and spin diffusion length. Our findings establish the purely spintronic nature of the aforementioned effects and provide a quantitative description in particular of the spin Seebeck effect. 
\end{abstract}
\maketitle

Pure spin currents present a new paradigm in spintronics~\cite{Wolf:2001, Zutic:2004} and spin caloritronics~\cite{Bauer:2012}. In particular, spin currents are the origin of spin pumping~\cite{Tserkovnyak:2002,Mosendz:2010}, the spin Seebeck effect~\cite{Uchida:2008,Uchida:2010a} and the spin Hall magnetoresistance (SMR)~\cite{Nakayama:2012, Althammer:2013, Vliestra:2013}. Taken alone, all these effects have been extensively studied, both experimentally~\cite{Czeschka:2011, Uchida:2008,Uchida:2010a, Nakayama:2012, Althammer:2013,	Hahn:2013,Rezende:2013} and theoretically~\cite{Tserkovnyak:2002, Xiao:2010,*Xiao_erratum:2010,Jiao:2012,Adachi:2011,Chen:2013,Tikhonov:2013}. 
From a theoretical point of view, all these effects are governed by the generation of a current of angular momentum via a non-equilibrium process. The flow of this spin current across a ferromagnet/normal metal interface can then be detected. The relevant interface property that determines the spin current transport thereby is the spin mixing conductance. Nevertheless, there has been an ongoing debate regarding the physical origin of the measurement data acquired in spin Seebeck and SMR experiments due to possible contamination with anomalous Nernst effect~\cite{Nernst:1887, Miyasato:2007,Huang:2011} or anisotropic magnetoresistance~\cite{Thomson:1856, Huang:2012} caused by static proximity polarization of the normal metal~\cite{Huang:2012}. To settle this issue, a rigorous check of the consistency of the spin-current based physical models across all three effects is needed. If possible contamination effects are absent, according to the spin mixing conductance concept~\cite{Brataas:2000}, there should exist a generalized Ohm's law between the interfacial spin current and the energy associated with the corresponding non-equilibrium process. This relation should invariably hold for the spin pumping, spin Seebeck and spin Hall magnetoresistance effects, as they are all based on the generation and detection of interfacial, non-equilibrium spin currents. We here put forward heuristic arguments that are strongly supported by experimental evidence for a scaling law that links all aforementioned spin(calori)tronic effects on a fundamental level and allows to trace back their origin to pure spin currents.


\begin{figure}
  \centering
  \includegraphics{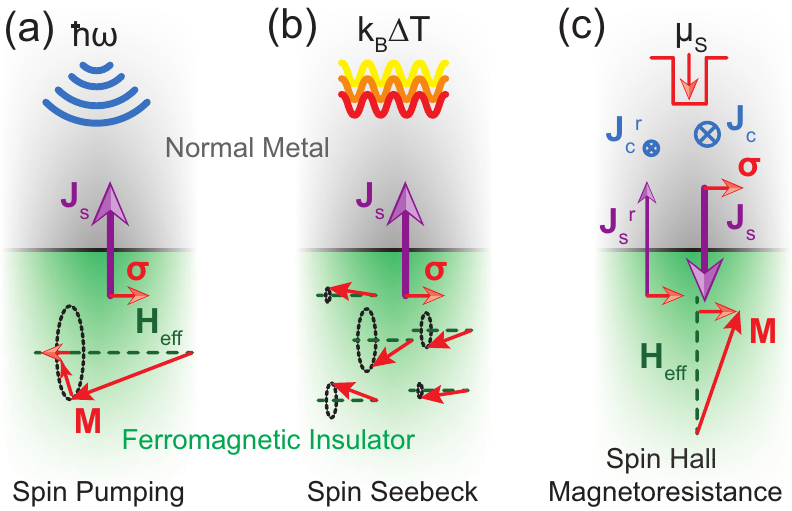}\\
  \caption{(color online) (a), (b) Schematic depictions of the spin pumping and spin Seebeck effects. The magnetization $\bm{M}$ in the ferromagnet (YIG in our experiments) is excited either resonantly (a) or thermally (b). The $\bm{M}$ precession around $\bm{H}_\mathrm{eff}$ is damped via the emission of a spin current $\bm{J}_\mathrm{s}$ with polarization $\bm{\sigma}$ into the normal metal (Pt in our experiments). (c) The spin Hall magnetoresistance is due to the torque exerted on $\bm{M}$ by an appropriately polarized $\bm{J}_\mathrm{s}$ which yields a change in the reflected spin current $\bm{J}_\mathrm{s}^\mathrm{r}$. The interconversion between $\bm{J}_\mathrm{s}$ ($\bm{J}_\mathrm{s}^\mathrm{r}$) and the charge currents $\bm{J}_\mathrm{c}$ ($\bm{J}_\mathrm{c}^\mathrm{r}$) are due to the (inverse) spin Hall effect in the normal metal.}\label{fig:schematics}
\end{figure}
%


We carried out a systematic set of spin pumping, spin Seebeck and SMR experiments on Y$_3$Fe$_5$O$_{12}$ (YIG) / Pt thin film bilayers. In our spin pumping experiments [schematically depicted in Fig.~\ref{fig:schematics}(a)], we place YIG / Pt bilayers in a microwave cavity operated at $\nu=\unit{9.85}{\giga\hertz}$ to resonantly excite magnetization dynamics. The emission of a spin current density $\bm{J}_\mathrm{s}$ across the bilayer interface into the Pt provides a damping channel for the non-equilibrium excitations of the magnetization $\bm{M}$. It has been established that the magnitude of the DC spin current density is given by~\cite{Tserkovnyak:2002}
\begin{equation}\label{eq:jssp}
\Jssp=\frac{\gSpinMix}{2\pi} \frac{1}{2} h \nu P \sin^2\Theta \;,
\end{equation}
where $\nu$ is the frequency of the microwave, $\Theta$ is the cone angle which the precessing magnetization $\bm{M}$ encloses with the effective magnetic field $\bm{H}_\mathrm{eff}$, $h$ is the Planck constant, $P$ is a factor to correct for elliptical precession of $\bm{M}$~\cite{Ando:2009} and \gSpinMix is the spin mixing conductance per unit of interface area and the conductance quantum $e^2/h$.

As shown in Fig.~\ref{fig:schematics}(b), thermal excitations of $\bm{M}$ also give rise to a spin current. This is the so-called spin Seebeck effect~\cite{Uchida:2008,Uchida:2010a}. Given a temperature difference $\Delta T$ between the electrons in the normal metal and the magnons in the ferromagnet, a DC spin current density~\cite{Xiao:2010} 
\begin{equation}\label{eq:jssse}
\Jssse=\frac{\gSpinMix}{2\pi} \frac{\gamma \hbar}{M_\mathrm{s} V_\mathrm{a}} k_\mathrm{B} \Delta T\;,
\end{equation}
is generated. We investigate the longitudinal spin Seebeck effect~\cite{Uchida:2010}, where the temperature gradient is applied across the ferromagnetic insulator/normal metal interface. In Eq.~\eqref{eq:jssse}, $\gamma=g \mu_\mathrm{B} / \hbar$ is the gyromagnetic ratio with the effective g-factor $g$ and the Bohr magneton $\mu_\mathrm{B}$, $M_\mathrm{s}$ is the saturation magnetization and $V_\mathrm{a}$ is the magnetic coherence volume given by~\cite{Xiao:2010,*Xiao_erratum:2010}
\begin{equation}\label{eq:Va}
V_\mathrm{a}= \frac{2}{3\zeta(5/2)}\left(\frac{4\pi D}{k_\mathrm{B} T}\right)^{3/2}\;, 
\end{equation}
with the Riemann Zeta function $\zeta$, the spin wave stiffness $D$ and $T=\unit{300}{\kelvin}$ for our room temperature experiments.

As depicted in Fig.~\ref{fig:schematics}(c), the application of a dc charge current density $\bm{J}_\mathrm{c}$ furthermore allows to inject a dc spin current density direction vector $\bm{J}_\mathrm{s}\propto\alpha_\mathrm{SH}\bm{J}_\mathrm{c}\times\bm{\sigma}$ into the YIG via the spin Hall effect in Pt~\cite{Nakayama:2012}. Here, $\alpha_\mathrm{SH}$ is the spin Hall angle of Pt and  $\bm{\sigma}$ is the spin current polarization.
If the magnetization $\bm{M}$ of the ferromagnet is oriented perpendicular to $\bm{\sigma}$, $\bm{J}_\mathrm{s}$ can exert a torque on $\bm{M}$ by being absorbed at the interface. When $\bm{\sigma}$ is parallel to $\bm{M}$, the spin current is reflected at the interface, causing a  spin current $\bm{J}_\mathrm{s}^\mathrm{r}$. Due to the inverse spin Hall effect, $\bm{J}_\mathrm{s}^\mathrm{r}$ again generates a charge current density $\bm{J}_\mathrm{c}^\mathrm{r}\propto\alpha_\mathrm{SH}\bm{J}_\mathrm{s}^\mathrm{r}\times\bm{\sigma}$ that effectively changes the electrical resistance of the Pt film.

The net spin current density $J_\mathrm{s}^\mathrm{SMR}=J_\mathrm{s}-J_\mathrm{s}^\mathrm{r}$ for $\bm{M}\parallel\bm{J}_\mathrm{c}$ is given by~\cite{Chen:2013}
\begin{equation}\label{eq:jssmr}
\Jssmr = \frac{\gSpinMix}{2\pi}  2 e \lambda_\mathrm{SD} \rho_\mathrm{Pt} \alpha_\mathrm{SH} J_\mathrm{c} \tanh \frac{t_\mathrm{Pt}}{2 \lambda_\mathrm{SD}} \eta \;,
\end{equation}
where $e$ is the elementary charge, $\lambda_\mathrm{SD}$ is the spin diffusion length in Pt and $\rho_\mathrm{Pt}$ and $t_\mathrm{Pt}$ are Pt resistivity and thickness, respectively. We furthermore introduced the correction factor~\cite{Chen:2013, Jiao:2012}
%
%
\begin{equation}\label{eq:eta}
\eta=\left[1+2 \gSpinMix \rho_\mathrm{Pt} \lambdaSD \frac{e^2}{h}  \coth \frac{t_\mathrm{Pt}}{\lambdaSD}\right]^{-1}\;.
\end{equation}

As suggested by Eqs.~\eqref{eq:jssp}, ~\eqref{eq:jssse} and~\eqref{eq:jssmr}, one should thus observe a scaling $J_\mathrm{s}=\frac{\gSpinMix}{2 \pi} E$ with an appropriate energy $E^\mathrm{SP}$, $E^\mathrm{SSE}$ and $E^\mathrm{SMR}$ that generates the spin pumping, spin Seebeck and SMR effects, respectively.  Note that, due to the inclusion of spin backflow via the correction factor $\eta$, the spin mixing conductance enters the linear response expression in a non-linear fashion by defining an effective excitation energy.

To experimentally test the scaling between $J_\mathrm{s}$ and $E$, we performed a series of spin pumping, spin Seebeck and spin Hall magnetoresistance measurements on more than 20 samples. Most of these samples consisted of thin film YIG / Pt bilayers with varying Pt thickness. Additionally, we used YIG / X / Pt trilayers in which X was a normal metal (Au or Cu). A complete list of samples, details of their preparation and relevant material parameters can be found in the appendix.

\begin{figure}
  \centering
  \includegraphics{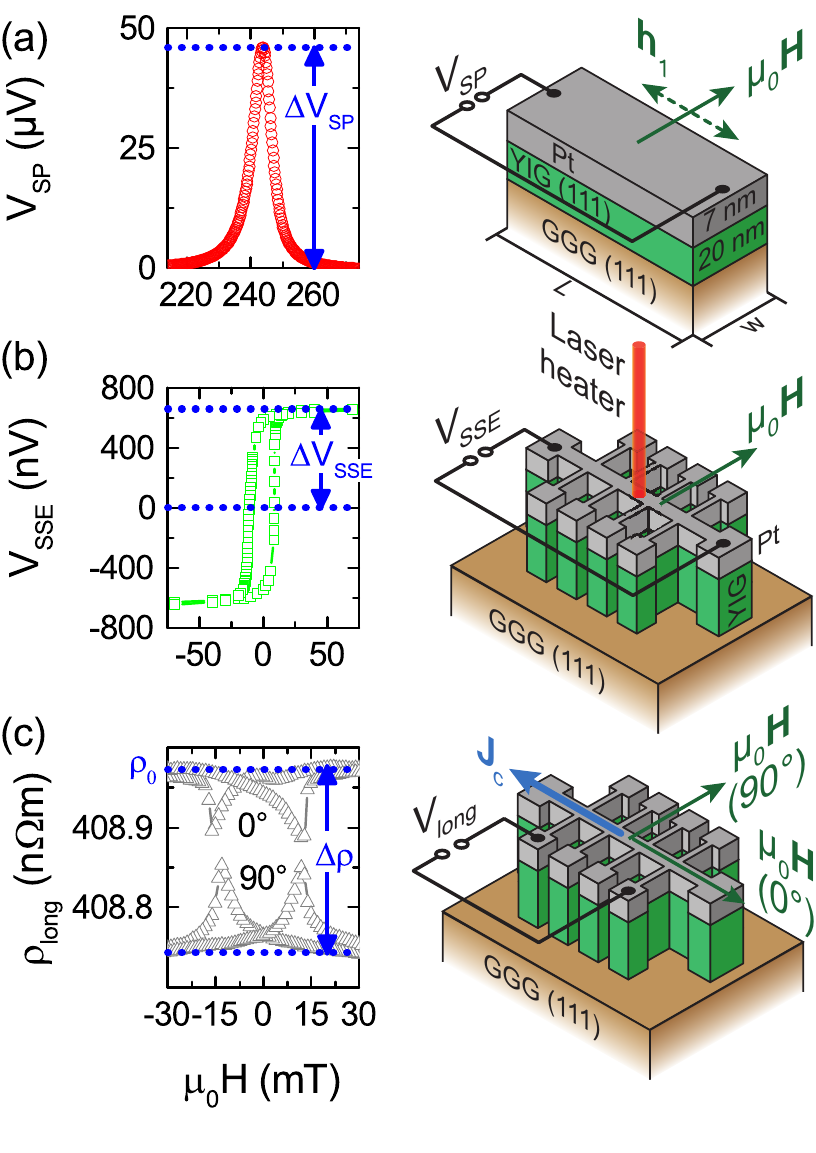}\\
  \caption{(color online) (a) Typical spin pumping data obtained from a YIG(20 nm)/ Pt (7 nm) bilayer sample as sketched to the right. $\Delta V_\mathrm{SP}$ is extracted in ferromagnetic resonance as indicated. (b) Data from a spin Seebeck experiment performed using a piece of the same sample. A laser beam is used to generate the thermal perturbation (see text) and $\Delta V_\mathrm{SSE}$ is obtained by taking half of the voltage difference observed between positive and negative saturation magnetic fields in the geometry sketched to the right. (c) DC magnetoresistance measurements are used to extract $\Delta \rho / \rho_0$ as the change in bilayer resistance upon rotating $\bm{M}$ from parallel (0\degree) to perpendicular (90\degree) to $\bm{J}_\mathrm{c}$. }\label{fig:exp}
\end{figure}
In spin pumping experiments with electrical spin current detection via the inverse spin Hall effect~\cite{Hirsch:1999, Ando:2011}, it is possible to determine \Jssp from the recorded dc voltage $\Delta V_\mathrm{SP}$ in ferromagnetic resonance (FMR) as~\cite{Mosendz:2010, Czeschka:2011}
\begin{equation}\label{eq:Jssp_Vdc}
\Jssp= \frac{\Delta V_\mathrm{SP}}{C \eta} \frac{1}{L} \;,
\end{equation}
with the sample length $L$ and the open-circuit spin Hall conversion efficiency~\cite{Czeschka:2011}
\begin{equation}\label{eq:C}
C=\frac{2e}{\hbar} \alphaSH \lambdaSD \tanh\left( \frac{t_\mathrm{Pt}}{2 \lambdaSD}\right) \frac{\rho_\mathrm{Pt}}{t_\mathrm{Pt}}\;,
\end{equation}
with $\hbar=h/(2 \pi)$. The factor $\eta$ is again included to correct for spin backflow~\footnote{Strictly speaking, for the SMR, the correction factor $\eta$ is not a backflow correction but is only obtained in a method similar to the backflow correction for spin pumping}. Note that Eq.~\eqref{eq:eta}, is strictly valid only for $2\pi\nu \tau_\mathrm{sf} \ll 1$ with the spin flip time $\tau_\mathrm{sf}\approx \unit{0.01}{\pico\second}$ in Pt~\cite{Jiao:2012}. $\eta$ is thus exact for the dc SMR and a good approximation for our spin pumping data ($2\pi\nu \tau_\mathrm{sf}\approx 6\times10^{-4}$).

A typical experimental $V_\mathrm{SP}$ trace recorded for a YIG (20 nm)/ Pt (7 nm) sample at a fixed microwave frequency $\nu=\unit{9.85}{\giga\hertz}$ while sweeping the external magnetic field $H$ is shown together with a schematic of the sample in Fig.~\ref{fig:exp}(a). We observe a resonant Lorentzian line shape of $V_\mathrm{SP}$ as a function of the external magnetic field~\cite{Czeschka:2011}.
Within experimental error, $V_\mathrm{SP}=0$ far away from FMR and $\Delta V_\mathrm{SP}$ is the dc voltage recorded at the resonance magnetic field as indicated in Fig.~\ref{fig:exp}(a).  
Because YIG is a ferrimagnetic insulator and we took great care to position the sample in a node of the microwave electric field, $\Delta V_\mathrm{SP}$ is not contaminated with rectification voltages~\cite{Bai:2008}. This is supported by the purely symmetric Lorentzian resonance line shapes observed. The sample length $L$ ranged from $\unit{3}{\milli\meter}$ to $\unit{5}{\milli\meter}$ in the different samples investigated. We determined $\rho_\mathrm{Pt}$ from four point resistance measurements and $t_\mathrm{Pt}$ from X-ray reflectometry. 

We now turn to the evaluation of $E^\mathrm{SP}=\frac{1}{2} h \nu P \sin^2 \Theta$ [cf.~Eq.~\eqref{eq:jssp}]. To this end, we extract $\Theta=2h_\mathrm{MW}/\Delta H$~\cite{Guan:2007} from the FWHM line width of the $\Delta V_\mathrm{SP}$ traces, where $\mu_0 h_\mathrm{MW}=\unit{22}{\micro\tesla}$ is the circular microwave magnetic field that was  determined from paramagnetic resonance calibration. We find $0.08\degree \leq \Theta \leq 0.55\degree$ in the different samples investigated. We calculate the ellipticity correction factor $P=1.2$ as detailed in Refs.~\cite{Mosendz:2010,Ando:2009} using a saturation magnetization $M_\mathrm{s}=\unit{140}{\kilo\ampere\per\meter}$ and an effective g factor $g=2$~\cite{Dorsey:1993}. We are now able to evaluate \Jssp and $E^\mathrm{SP}$ as a function of the three parameters $\gSpinMix$, $\alpha_\mathrm{SH}$ and $\lambda_\mathrm{SD}$. We discuss below that with a single set of these parameters we can quantitatively describe the spin pumping, spin Seebeck and SMR data in the context of the spin mixing conductance concept.

In a different set of experiments, using parts of the same samples patterned into Hall bar mesas by optical lithography and subsequent Ar-ion etching, we determined the dc voltage $\Delta V_\mathrm{SSE}$ due to the laser-heating induced spin Seebeck effect~\cite{Weiler:2012}. A laser beam of adjustable power ($\unit{1.8}{\milli\watt}\leq P_\mathrm{L}\leq\unit{57}{\milli\watt}$) impinges on the main Hall bar (length $L=\unit{950}{\micro\meter}$ and width $w=\unit{80}{\micro\meter}$) which is oriented perpendicular to the external, in-plane magnetic field. The laser beam is dominantly absorbed in the Pt layer and yields a temperature difference $\Delta T$ between the magnons in YIG and the electrons in the Pt at the YIG/Pt interface. We use a numerical model incorporating a thermal contact resistance between the YIG and the normal metal layers to compute the magnon, phonon and electron temperature profiles in our samples as a function of layer composition and laser power~\cite{Schreier:2013}.  We find $\unit{0.02}{\kelvin}\leq\Delta T\leq\unit{0.9}{\kelvin}$. The spin current \Jssse is detected via the inverse spin Hall voltage $V_\mathrm{SSE}$ along the main Hall bar. A typical $V_\mathrm{SSE}$ curve is shown in Fig.~\ref{fig:exp}(b) as a function of the external magnetic field. The depicted hysteretic $V_\mathrm{SSE}$ vs. $H$ loop is consistent with our previous experiments~\cite{Weiler:2012}. The spin current density is extracted from experiment by
\begin{equation}\label{eq:Jssse_Vdc}
\Jssse= \frac{\Delta V_\mathrm{SSE}}{C\eta}\frac{2w}{a^2 \pi}
\end{equation}
where $L$ from Eq.~\eqref{eq:Jssp_Vdc} is now replaced by $a^2 \pi / 2w$ with the laser spot radius $a=\unit{2.5}{\micro\meter}$ and the Hall bar width $w=\unit{80}{\micro\meter}$. This stems from lateral integration over the Gaussian laser spot profile to account for the fact that the sample is heated only locally as demonstrated in Ref.~\cite{Weiler:2012} and is valid as long as $a\ll w$, which is the case for all investigated samples. We use the identical values for $C$ and $\eta$ as for the spin pumping experiments to calculate $\Jssse$ with Eq.~\eqref{eq:Jssse_Vdc}. To quantify $E^\mathrm{SSE}=\gamma \hbar k_\mathrm{B} \Delta T/ (M_\mathrm{s} V_\mathrm{a})$, we use the coherence volume $V_\mathrm{a}=(\unit{1.3}{\nano\meter})^3$ which we obtain from Eq.~\eqref{eq:Va} by using $D=\unit{8.5\times10^{-40}}{\joule\meter\squared}$ consistent with theory and a broad range of experiments~\cite{Srivastava:1987}.

In yet further, independent experiments with the same set of Hall-bar samples we measured the SMR as the change $\Delta \rho$ in bilayer resistance $\rho$ when rotating $\bm{M}\parallel\bm{J}_\mathrm{c}$ ($\rho_0$) to $\bm{M}\perp \bm{J}_\mathrm{c}$ ($\rho_0+\Delta \rho$). Typical $\rho(H)$ traces for $\bm{M}\parallel\bm{J}_\mathrm{c}$ (0\degree) and $\bm{M}\perp\bm{J}_\mathrm{c}$ are shown in Figure~\ref{fig:exp}(c). One can observe the magnetization switching at the coercive magnetic fields that agree with those extracted from our spin Seebeck experiments on identical samples. As detailed in Ref.~\cite{Althammer:2013}, the change in $\rho$ with $H$ can be traced back to the SMR from its characteristic dependence on $\bm{M}$ orientation. From the SMR data, we extract the spin current density~\cite{Chen:2013}
\begin{equation}\label{eq:SMR_deltarho}
\Jssmr=J_\mathrm{c} \frac{\Delta \rho}{\rho_0} \frac{\hbar t_\mathrm{Pt}}{\alpha_\mathrm{SH} e \lambda_\mathrm{SD} \tanh \frac{t_\mathrm{Pt}}{2 \lambda_\mathrm{SD}}}
\end{equation}
from the experimentally determined $\Delta \rho / \rho_0$.  The charge current densities in our experiments were $\unit{1.7\times10^6}{\ampere\per\meter\squared}\leq J_\mathrm{c}\leq\unit{1.7\times10^9}{\ampere\per\meter\squared}$.

\begin{figure}
  \centering
  \includegraphics[]{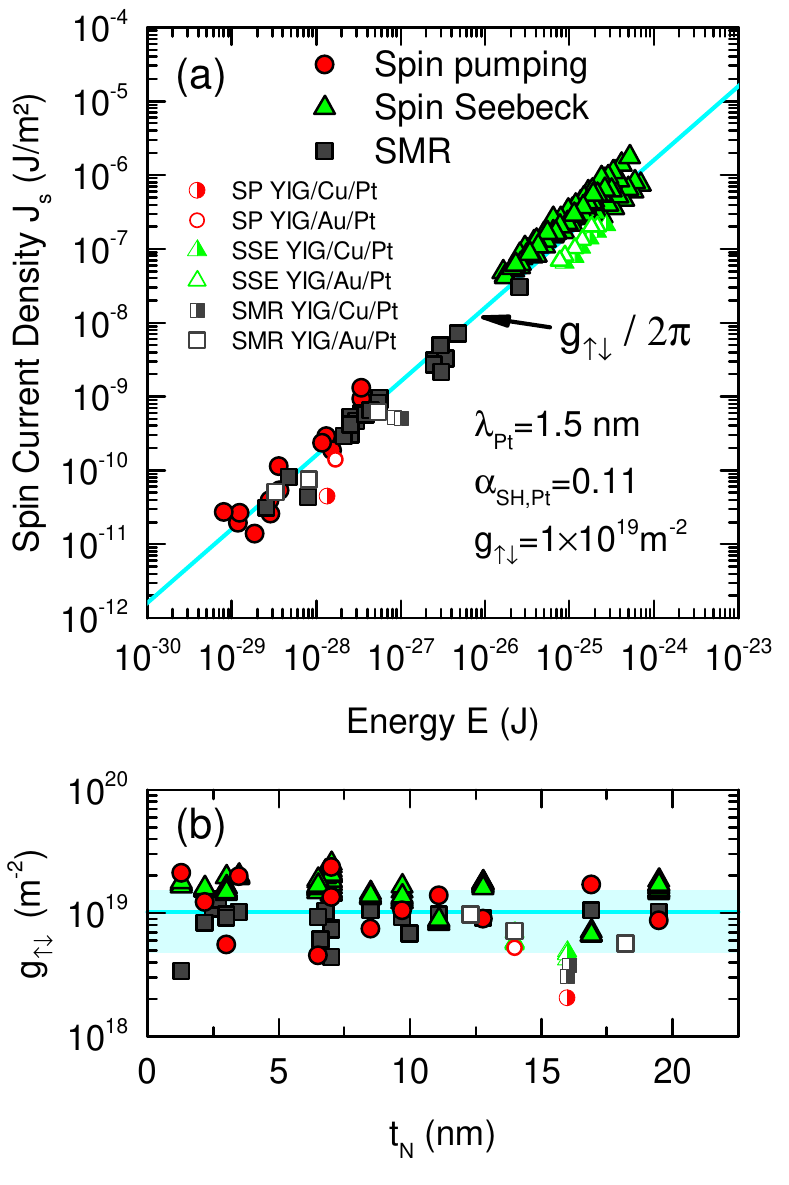}\\
  \caption{(color online) (a) Spin current density $J_\mathrm{s}$ as a function of the non-equilibrium energy $E$ for all investigated samples as determined in spin pumping (circles), spin Seebeck (triangles) and SMR (squares) measurements. The solid line is the proportionality constant identified in the text as $\gSpinMix / 2 \pi$.  Open symbols correspond to YIG/Au/Pt trilayer samples and half-filled symbols to YIG/Cu/Pt trilayer samples. (b) The spin mixing conductance as a function of total normal metal thickness [same symbols as in (a)].}\label{fig:scaling}
\end{figure}

We plot $J_\mathrm{s}^\mathrm{SMR}$ from Eq.~\eqref{eq:SMR_deltarho} as a function of $E^\mathrm{SMR}=2 e \lambda_\mathrm{SD} \rho_\mathrm{Pt} \alpha_\mathrm{SH} J_\mathrm{c} \tanh\left(t_\mathrm{Pt}/(2 \lambda_\mathrm{SD})\right) \eta $ in Fig.~\ref{fig:scaling}(a) (squares) for all samples. In identical fashion, Fig.~\ref{fig:scaling}(a) depicts $J^\mathrm{SP}$ as a function of $E^\mathrm{SP}$ (circles) and  $J^\mathrm{SSE}$ as a function of $E^\mathrm{SSE}$ (up triangles). We use a single set of parameters, $\gSpinMix=\unit{1\times10^{19}}{\meter^{-2}}$, $\alpha_\mathrm{SH}=0.11$ and $\lambda_\mathrm{SD}=\unit{1.5}{\nano\meter}$ for all samples. These parameters are identical to those extracted from an analysis of the Pt thickness dependence of the SMR~\cite{Althammer:2013}. We acquired data points for SMR and spin Seebeck effect on various samples as a function of charge current density or laser power, respectively.  We furthermore include data recorded using YIG/Au/Pt (open symbols) and YIG/Cu/Pt (half-filled symbols) trilayer samples (symbol shape identifies spin pumping, spin Seebeck or SMR data). To evaluate the trilayer data, we assume vanishing $\alpha_\mathrm{SH}$ and $\lambda_\mathrm{SD}\gg t$ for Au and Cu. We thus modify $C$ for spin pumping and spin Seebeck effect as well as $J_\mathrm{c}$ and $\frac{\Delta \rho}{\rho_0}$ for the SMR as detailed in the appendix. 

Altogether our experimental data span five orders of magnitude in $J_\mathrm{s}$ and $E$. In this entire range, we observe that all experimental data points fall on (or close to) one line in the plot. As predicted by theory, the constant of proportionality is found to be $\gSpinMix / 2 \pi$.  This has several implications. First, and most importantly, Fig.~\ref{fig:scaling}(a) is strong evidence for the spin mixing conductance concept, i.e., that spin pumping, spin Seebeck and SMR effects indeed arise from pure spin currents physics. Spurious effects due to static proximity polarization in Pt~\cite{Huang:2011, Huang:2012, Qu:2013} can be excluded based on Fig.~\ref{fig:scaling}(a), because if the measured spin Seebeck effect and spin Hall magnetoresistance data were indeed consequence of a static proximity polarization, the obtained data should not fit quantitatively in the spin current injection picture parameterized by the mixing conductance. Our data for both these effects however are consistent with those extracted from spin pumping experiments, showing that all three effects have one and the same microscopic origin, namely spin current flow. Last but not least, our data enable a quantitative understanding of the spin Seebeck effect which has remained elusive so far~\cite{Agrawal:2012}.  The scaling of our data shows that a spin Hall angle $\alpha_\mathrm{SH}=0.11$ and spin diffusion length $\lambda_\mathrm{SD}=\unit{1.5}{\nano\meter}$ are reasonable material parameters for our Pt thin films. These parameters are in accordance with the product $\alpha_\mathrm{SH}\lambda_\mathrm{SD}$ extracted in previous studies~\cite{Vila:2007, Mosendz:2010} and more recent findings~\cite{Liu:2011, Castel:2012, Boone:2013}. In our work, owing to the different functional dependence of the three effects on these parameters, a more reliable extraction of $\lambda_\mathrm{SD}$ and $\alpha_\mathrm{SH}$ becomes possible. 

Figure~\ref{fig:scaling}(b) shows $\gSpinMix= 2\pi J_\mathrm{s} /E$ as a function of the total normal metal thickness $t_\mathrm{N}=t_\mathrm{Pt}+t_\mathrm{X}$ (X is Au or Cu) for all bilayer and trilayer samples. The symbol definitions are identical to that in Fig.~\ref{fig:scaling}(a). The solid line depicts $\gSpinMix=\unit{1\times10^{19}}{\meter^{-2}}$  and the shaded region corresponds to $\unit{0.5\times10^{19}}{\meter^{-2}}\leq\gSpinMix\leq\unit{1.5\times10^{19}}{\meter^{-2}}$. The majority of our data points lie within the shaded region, so \gSpinMix is constant within $\pm50\%$ for all our samples and regardless of the experimental method used to extract it. There is no discernible trend in the $J_\mathrm{s}$ to $E$ ratio as a function of Pt (or X/Pt) thickness. This suggests that Eqs.~\eqref{eq:Jssp_Vdc},~\eqref{eq:Jssse_Vdc} and~\eqref{eq:SMR_deltarho} are sufficiently accurate in the entire thickness range investigated. The unsystematic scatter in \gSpinMix in Fig.~\ref{fig:scaling}(b) can be accounted for as being experimental errors and varying interface properties between the different samples. Our trilayer samples exhibit a \gSpinMix  similar or slightly lower than that of our YIG/Pt bilayers and previous findings for the YIG/Au interface~\cite{Heinrich:2011, burrowes:2012}.

In summary, we have experimentally demonstrated that spin pumping, spin Seebeck and SMR all share the same purely spintronic origin and thus experimentally validated the spin mixing conductance concept. Spurious contributions due to proximity ferromagnetism in Pt can be ruled out, thereby supporting existing models for SMR and spin Seebeck effect. A relevant set of parameters for a ferromagnetic insulator/normal metal bilayer (or according trilayer) obtained from rather straightforward SMR experiments may be used to predict results for spin Seebeck or spin pumping experiments on the same samples.

Financial support from the DFG via SPP 1538 ''Spin Caloric Transport", Project No. GO 944/4-1, the Dutch FOM Foundation, EC Project ''Macalo", the National Natural Science Foundation of China (No. 11004036, No. 91121002) and the German Excellence Initiative via the ''Nanosystems Initiative Munich" (NIM) is gratefully acknowledged.

\section{Appendix}
\subsection{Spin pumping and spin Seebeck voltages}

\begin{figure}
\centering
\includegraphics{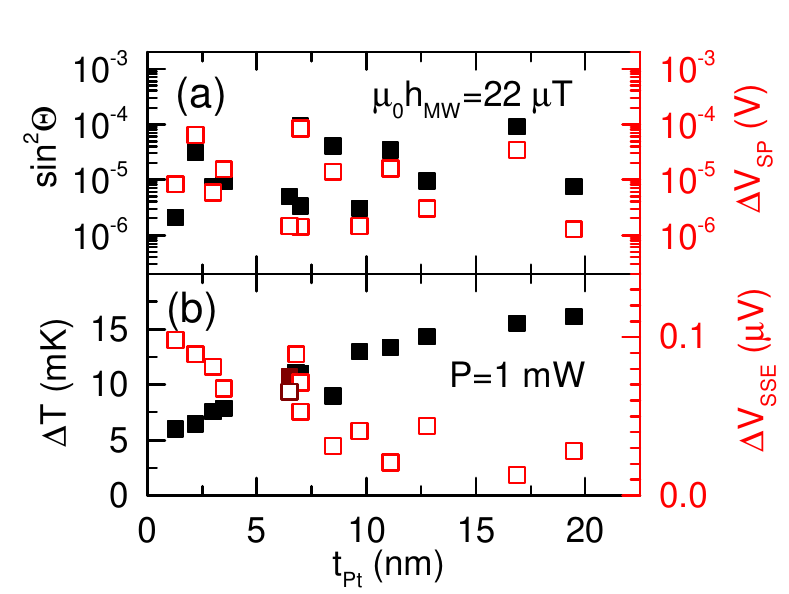}
\caption{(a) Precession cone angle $\Theta$ (left scale, closed symbols) and spin pumping voltage $\Delta V_\mathrm{SP}$ (right scale, open symbols) determined from experiments as a function of Pt thickness $t_\mathrm{N}$ for constant microwave driving field $\mu_0h_\mathrm{MW}=\unit{22}{\micro\tesla}$ at $\nu=\unit{9.85}{\giga\hertz}$.  (b) Calculated magnon-electron temperature difference $\Delta T$ (left scale, closed symbols) between YIG and Pt at the YIG/Pt interface and measured spin Seebeck voltage $\Delta V_\mathrm{SSE}$ (right scale, open symbols) at constant laser power $P=\unit{1}{\milli\watt}$ impinging on the sample. }\label{fig:SI_V}
\end{figure}

In this section, we present the experimentally acquired spin pumping and spin Seebeck voltages for our YIG/Pt bilayer samples. 

Figure~\ref{fig:SI_V}(a) shows the precession cone angle $\Theta$ as determined from the line width of the ferromagnetic resonance spectra and the spin pumping voltage $\Delta V_\mathrm{SP}$ determined from experiment for constant microwave driving field $\mu_0 h_\mathrm{MW}=\unit{22}{\micro\tesla}$ at $\nu=\unit{9.85}{\giga\hertz}$.  At constant $h_1$, $\Theta$ is determined mainly by the properties of the ferromagnet and thus there is no systematic evolution of $\Theta$ with $t_\mathrm{N}$. Generally however, it can be observed that larger $\Theta$ results in larger $\Delta V_\mathrm{SP}$ as already demonstrated in~\cite{Czeschka:2011}. Furthermore, it can be seen that due to electrical short circuiting, the inverse spin Hall effect conversion of the spin current to $\Delta V_\mathrm{SP}$ tends to be more efficient at small $t_\mathrm{N}$.

Figure~\ref{fig:SI_V}(b) depicts the simulated magnon-electron temperature difference $\Delta T$ between YIG and Pt at the YIG/Pt interface. For details of the calculations that rely on a finite Kapitza interface resistance refer to Ref.~\cite{Schreier:2013}. A systematic evolution of $\Delta T$ as a function of $t_\mathrm{N}$ is observed. The reduced $\Delta T$ at small $t_\mathrm{N}$ is due to the incomplete absorption of the laser light. The scatter in the simulated $\Delta T$ reflects the influence of spin backflow on $\Delta T$ which hence depends on the normal metal resistivity~\cite{Schreier:2013}. The observed spin Seebeck voltage $\Delta V_\mathrm{SSE}$ is reduced for large $t_\mathrm{N}$ due to electrical short circuiting. 

A detailed analysis of the spin Hall magnetoresistance properties of the samples used in this study can be found in Ref.~\cite{Althammer:2013}.

\subsection{Sample preparation}
All  Y$_3$Fe$_5$O$_{12}$ (YIG) thin films were grown (111) oriented on either Gd$_3$Ga$_5$O$_{12}$ (GGG) or Y$_3$Al$_5$O$_{12}$ (YAG) substrates by pulsed laser deposition and covered in situ with Au, Cu and/or Pt by electron beam evaporation. All samples were diced in two parts. One part was patterned into a Hall bar geometry for laser-induced longitudinal spin Seebeck and spin Hall magnetoresistance measurements while the unpatterned part was used for the spin pumping experiments. Table~\ref{table:bisamples} lists all investigated bilayer samples and Table~\ref{table:trisamples} the trilayer samples. 
\begin{table}
\begin{tabular}{lc}\toprule    
  Sample & $\rho_\mathrm{Pt}$ (n$\Omega$m) \\ 
\colrule    
GGG/YIG(50)/Pt(7) & 409.4 \\
GGG/YIG(54)/Pt(7) & 406.5 \\
GGG/YIG(53)/Pt(2.5) & 719 \\
GGG/YIG(65)/Pt(6.6) & 582.6 \\
GGG/YIG(46)/Pt(3.5) & 306.6 \\
GGG/YIG(69)/Pt(2.7) & 453.6 \\
GGG/YIG(58)/Pt(2.2) & 761.7 \\
GGG/YIG(57)/Pt(1.3) & 1089.9 \\
GGG/YIG(61)/Pt(11.1) & 334.5 \\
GGG/YIG(52)/Pt(16.9) & 339.2 \\
GGG/YIG(53)/Pt(8.5) & 348.3 \\
YAG/YIG(59)/Pt(6.8) & 487.7 \\
YAG/YIG(64)/Pt(3) & 622.2 \\
YAG/YIG(61)/Pt(19.5) & 361.3 \\
YAG/YIG(63)/Pt(6.5) & 412 \\
YAG/YIG(60)/Pt(9.7) & 429 \\
YAG/YIG(60)/Pt(12.8) & 434.9 \\
YAG/YIG(50)/Pt(3) & 513 \\
\botrule   
\end{tabular}
\caption{Bilayer samples used in this study. Numbers in parentheses indicate layer thickness in nm.}
\label{table:bisamples}
\end{table}
\begin{table}
\begin{tabular}{lc}\toprule    
  Sample & $\rho_\mathrm{Pt}$ (n$\Omega$m)\\ 
  \colrule    
GGG/YIG(31)/Cu(8.8)/Pt(7.3) & 410 \\
GGG/YIG(20)/Au(7)/Pt(7) & 400 \\
GGG/YIG(20)/Cu(9)/Pt(7) & 400 \\
YAG/YIG(55)/Au(9.2)/Pt(9) & 370 \\
YAG/YIG(45)/Au(9.4)/Pt(2.9) & 860 \\
\botrule   
\end{tabular}
\caption{Trilayer samples used in this study. We assumed $\rho_\mathrm{Cu}=\rho_\mathrm{Au}=\unit{300}{\nano\ohm\meter}$ for all trilayer samples.}
\label{table:trisamples}
\end{table}

\subsection{Trilayers}
The trilayer spin pumping and spin Seebeck data are evaluated by using 
\begin{equation}
C_\mathrm{TL}=\frac{2e}{\hbar} \alphaSH \lambdaSD \tanh\left( \frac{t_\mathrm{Pt}}{2 \lambdaSD}\right) \frac{1}{t_\mathrm{Pt}/\rho_\mathrm{Pt}+t_\mathrm{X}/\rho_\mathrm{X}}\;,
\end{equation}
instead of $C$ in Eqs.~(6) and~(8) in the main text to account for the additional electrical shortcircuit due to X. For the spin Seebeck effect, we calculate $\Delta T$ as the magnon-electron temperature difference between the YIG and the adjacent normal metal for the trilayer samples.
For the SMR trilayer data, we again assume that only Pt contributes to the SMR, while the second metallic layer acts as a mere electrical short-circuit. We extract the charge current in Pt
\begin{equation}
J_\mathrm{c, Pt}=J_\mathrm{c} \frac{t_\mathrm{Pt} \rho_\mathrm{X}}{t_\mathrm{Pt} \rho_\mathrm{X}+t_\mathrm{X} \rho_\mathrm{Pt}}\;,
\end{equation}
by assuming that Pt and X layers act as parallel resistors for the charge current. We then derive the SMR in Pt
\begin{equation}
\frac{\Delta \rho_\mathrm{Pt}}{\rho_{0, \mathrm{Pt}}}=\frac{\Delta \rho}{\rho_0} \frac{t_\mathrm{Pt} \rho_\mathrm{X}+t_\mathrm{X} \rho_\mathrm{Pt}}{t_\mathrm{Pt} \rho_\mathrm{X}}\;,
\end{equation}
by assuming there is no change in $\rho_\mathrm{X}$ due to the spin Hall magnetoresistance. We use these expression to replace $J_\mathrm{c}$ in Eq.~(4) and $\frac{\Delta \rho}{\rho_0}$ in Eq.~(9) in the main text to evaluate $J_\mathrm{s}^\mathrm{SMR}$ and $E^\mathrm{SMR}$ for the trilayer samples.

\end{document}